\documentclass[prl,12pt,preprint]{revtex4}
\usepackage{amsfonts}
\usepackage{amsmath}
\usepackage{amssymb}
\usepackage{graphicx}

\begin{document}

\title{Bose condensation of interacting gases in traps with and without
optical lattice}
\author{S. Chatterjee, A. E. Meyerovich}
\affiliation{Department of Physics, University of Rhode Island, Kingston, RI 02881 - 0817}
\keywords{Bose condensation; trapped Bose gas;interacting bosons;harmonic
trap; optical lattice}
\pacs{03.75.Hh, 05.30.Jp, 67.85.Hj, 67.85.-d}

\begin{abstract}
We discuss effects of particle interaction on Bose condensation in
inhomogeneous traps with and without optical lattice. Interaction pushes
normal particles away from the condensate droplet, which is located in the
center of the trap, towards the periphery of the trap where the trapping
potential is large. In the end, the remaining normal particles are squeezed
to a quasi-$2D$ shell around the condensate droplet thus changing the
effective dimensionality of the system. In the absence of the optical
lattice the index in the temperature dependence of the condensate density at
the later stages of the process is close to 2 with a weak dependence on the
number of trapped particles. In the presence of the lattice inside the trap
this index acquires a strong dependence on the number of particles inside
the trap and gradually falls from a $3D$ to a $2D$ value with an increase in
the number of particles. This change in index is explained by the
lattice-driven spread of the condensate droplet and the localization of the
narrow band particles by the trap potential.
\end{abstract}

\startpage{1}
\maketitle

The spectacular experimental discovery of Bose condensation (BEC) made the
study of alkali gases in traps the focal point in atomic, low temperature,
and condensed matter physics. For the first time, it became possible to
observe some of the phenomena that have been discussed earlier only within
theoretical models (see review \cite{cor1}). The phenomena in ultracold
alkali gases are incredibly rich and combine features inherent to diverse
condensed matter and low temperature systems (Refs. \cite{review2} and
references therein) from "classical" superfluid or superconducting systems
\cite{review2} to spin-polarized quantum gases \cite{jila1} to Mott
transition in the optical lattice \cite{nature1}.

The trap potential is always inhomogeneous. The interplay between the
repulsive interaction and the trapping potential complicates BEC \cite%
{baym,pit} since these two factors have opposite effects on condensation:
while the trap tends to concentrate the condensate in a narrow region of
space around the particle ground state in the trap, the repulsion is
responsible for the widening of this condensate droplet. The analytical
description of the combined effects tends to be rather elusive and our
previous experience with condensation in homogeneous systems is not very
helpful. The problem becomes even more complex in the presence of the
optical lattice inside the trap which adds two different localization
processes - Mott transition and localization of narrow band particles by an
inhomogeneous potential.

Below we investigate a situation in which it is possible to get an accurate
picture of the condensation in trapped interacting gases. The main attention
is paid to the index in the temperature dependence of the condensate
fraction and to the size of the condensate droplet. It turns out that this
index is not universal even for a low density gas. What is more, the
effective dimensionality of the problem changes with condensation and the
later stages of BEC are different from initial.

We start from BEC in trapped gases without the optical lattice, and add the
complications associated with the optical lattice later on. We assume that
the density is sufficiently low to neglect the interaction before the onset
of condensation even in the center of the trap. This means that $T_{c}$ is
unaffected by the interaction. The interaction is brought into play only
with the onset of condensation since the particles condensate in the center
of the trap making the density in the center large. This makes the
interaction, which is proportional to the particle density, large only in
and around the condensate droplet. This also means that the normal particles
are pushed out by the dense condensate towards the periphery of the trap
where the interaction is negligible. The further particles move away from
the center the higher is the gradient of the trapping potential which is
responsible for the force pushing the normal particles back towards the trap
center. Thus, at the \emph{later} stages of BEC, the majority of remaining
normal particles are located in an almost two-dimensional shell around the
condensate droplet and the dimensionality of the problem changes from the
three-dimensional in the beginning of the condensation to quasi-$2D$ later
on.

We consider a $3D$ harmonic trap with a single-particle ground state of
frequency $\omega $ and spacial size $\sigma _{0}$ (axial asymmetry of real
traps is irrelevant in our context). Without interaction, BEC starts at $%
T_{c}=0.941\hbar \omega N^{1/3}$ \cite{fetter} and the size of the
condensate droplet is $\sigma _{0}$. Repulsion increases the size of the
droplet with $N_{c}\left( T\right) $ particles to $\sigma \left( T\right) $.
Then the potential well for normal particles $U\left( r\right) $ has a
shell-type structure,
\begin{equation}
U\left( r,T\right) =\frac{1}{2}\hbar \omega \left[ \frac{r^{2}}{\sigma
_{0}^{2}}+\frac{N_{c}\sigma _{0}^{3}}{N_{0}\sigma ^{3}}\exp \left(
-r^{2}/\sigma ^{2}\right) \right] ,  \label{eq1}
\end{equation}%
where $N_{0}=\left( \sqrt{\pi }/8\right) \omega m\sigma _{0}^{3}/\hbar a_{s}$
and we assume that the condensate density is Gaussian. The number of normal
particles $N_{n}\left( T\right) =N-N_{c}\left( T\right) $ is determined from
the condition $\mu =0$. The size of the condensate droplet $\sigma \left(
T\right) $ can be obtained from minimization of the condensate energy,
including repulsion, similarly to Ref. \cite{baym}. The interaction between
the normal particles can often be excluded from Eq. $\left( \ref{eq1}\right)
$. First, for less than $10^{5}$ particles in a trap, the density of the
normal particles is negligible even in the trap center. For larger $N$, the
number of the normal particles on the later stages of the condensation is
small. Finally, the density of the normal particles is suppressed even more
by repulsion from the condensate droplet which spreads them through a large
shell around the droplet $4\pi \sigma ^{2}\sigma _{0}$\ instead of
concentrating them near the center in the volume $\left( 4\pi /3\right)
\sigma _{0}^{3}$. This gives $N$ at least an extra order of magnitude for
which we can neglect the interaction of normal particles.

$N_{0}$ in Eq. $\left( \ref{eq1}\right) $ is the minimal number of particles
in the condensate that is sufficient to create a strong repulsive core in
the center of the trap. When $N>N_{c}\gg N_{0}$ the normal particles are
pushed away from the center by the repulsive core $\left( \ref{eq1}\right) $
into a potential valley surrounding the condensate droplet. For \textrm{Rb}
in a trap with $\omega =24$ \textrm{Hz}, the values $a_{s}=58.2$ $\mathrm{%
\mathring{A}}$, $\sigma _{0}=2.2\times 10^{-6}$ \textrm{m}, and the critical
number $N_{0}$ that changes the topology of the normal cloud is $%
N_{0}\approx 84$. The center of the trap becomes inaccessible for normal
particles when $T$ is much smaller than the repulsion from the core. Using $%
T_{c}$ instead of $T$ and $N$ instead of $N_{c}$, one gets $\sigma
^{3}N_{0}/\sigma _{0}^{3}\ll N^{2/3}$ and the critical value of $N_{c}$ is
around $10^{5}$. All this means that our results are applicable for $N$ in
the range $10^{4}\div 10^{6}$.

We are able to obtain an analytical description of the situation (\textit{cf.%
} Ref. \cite{pit}). At the later stages of the condensation, the potential $%
\left( \ref{eq1}\right) $ forms a distinct valley away from the center of
the trap as soon as $N_{c}\gg N_{0}$ and equations for $N_{c}\left( T\right)
$ and $\sigma \left( T\right) $ reduce to
\begin{eqnarray}
\chi  &=&\sqrt{2}-\frac{\sigma _{0}^{4}}{\sqrt{2}\sigma ^{4}}=\frac{%
4a_{s}\sigma _{0}^{4}}{\sqrt{\pi }\sigma ^{5}}N_{c},  \label{eq3} \\
N_{c} &=&N-\sum_{n=1}\left[ \exp \left[ \widetilde{\beta }\left( n+\frac{1}{2%
}\right) \lambda \right] -1\right] ^{-1}  \notag \\
&&-\sum_{n,l=0}\frac{2l+1}{\exp \widetilde{\beta }\left[ \left( n+\frac{1}{2}%
\right) \lambda +\left( l^{2}+l\right) \frac{\sigma _{0}^{2}}{2\sigma ^{2}}%
\ln 2\chi \right] -1}  \notag
\end{eqnarray}%
with $\widetilde{\beta }=\hbar \omega /T$, $\lambda =\sqrt{2\ln \left( 2\chi
\right) }$. The summation provides the temperature dependencies $N_{c}\left(
T\right) $ and $\sigma \left( T\right) $.

We found that the condensate fraction at the later stages of condensation
can be given as
\begin{equation}
N_{c}/N=1-\left( T/T_{c}^{\ast }\right) ^{\alpha }  \label{eq2}
\end{equation}%
with a relatively high accuracy. The important feature of Eq.$\left( \ref%
{eq2}\right) $ is that the temperature is normalized not by the critical
temperature $T_{c}$ for the onset of condensation but by a different value $%
T_{c}^{\ast }$. Since the squeezing of the normal particles towards the
fringes of the trap accelerates with the number of particles in the
condensate $N_{c}$, the normal shell narrows with increasing $N_{c}$, and,
therefore, $N$. As a result, the effective temperature $T_{c}^{\ast }$
should be higher than $T_{c}$ and increase with increasing $N$. Dependence
of $T_{c}^{\ast }$, or, more precisely, $T_{c}^{\ast }/\hbar \omega N^{1/3}$%
, on $N$ is presented in Figure 1. For comparison, the critical temperature $%
T_{c}$ for non-interacting particles in a $3D$ harmonic trap is $%
T_{c}=0.9\hbar \omega N^{1/3}$ \cite{fetter}.

\begin{figure}[ptb]
\includegraphics[angle=270,scale=0.5]{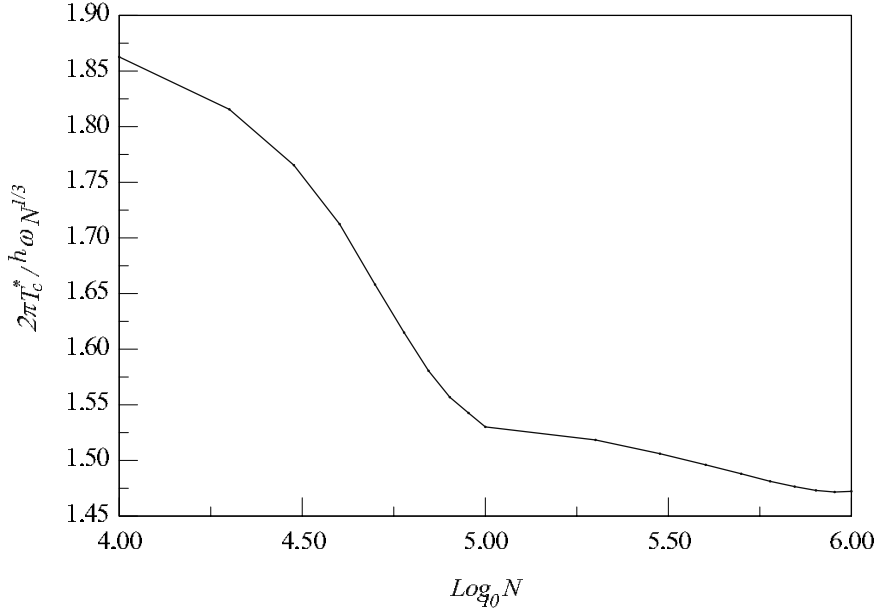}
\caption{Density dependence of the reduced critical temperature $%
T_{c}^{\ast }/\hbar \omega N^{1/3}$, Eq.$\left( \ref{eq2}\right) $. For a
non-interacting gas in a $3D$ harmonic trap this ratio should be 0.91.}
\label{fig1}
\end{figure}

The striking change in behavior of $T_{c}^{\ast }\left( N\right) $ in Figure
1 occurs at $N$ for which $T_{c}\sim \frac{1}{2}\hbar \omega \left(
N_{c}\sigma _{0}^{3}/N_{0}\sigma ^{3}\right) $. At higher densities the
repulsion from the condensate droplet keeps the normal particles near the
bottom of the potential valley around the droplet; at lower densities, the
normal particles spread out and can even reach the center of the trap. An
anomaly at the same threshold density is also observed in $\alpha \left(
N\right) $, Figure 2, though the index $\alpha $ remains very close to the
value $2$ and is practically independent of $N$, $\alpha =2.02\pm 1\%,$\ in
a wide range of $N$ from $10^{4}$ to $10^{6}$. This weak dependence $\alpha
\left( N\right) $ is surprising for a nonlinear problem of this nature. The
residual temperature dependence $\alpha \left( T\right) $ is within the same
error bars.

\begin{figure}[ptb]
\includegraphics[angle=270,scale=0.5]{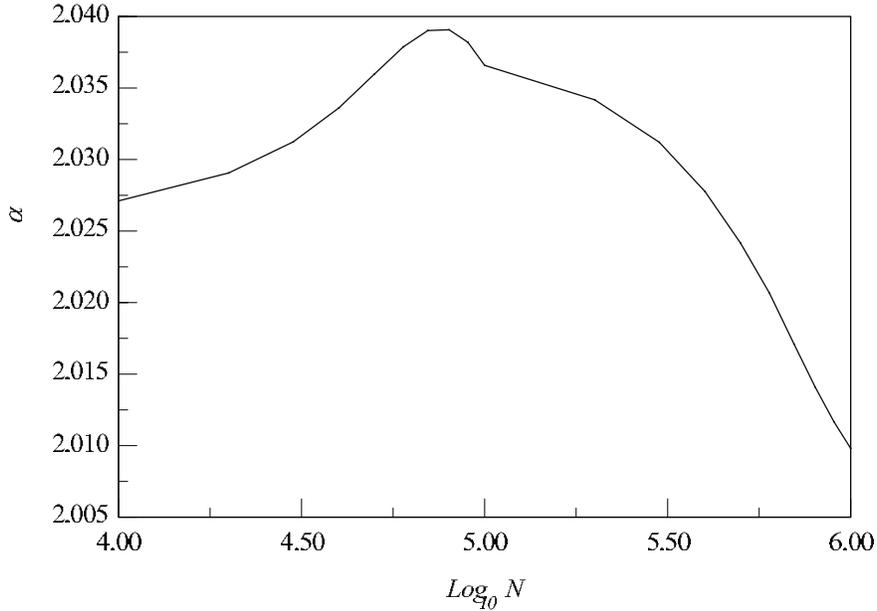}
\caption{Density dependence of the index $\alpha $, Eq.$\left( \ref{eq2}%
\right) .$ For a non-interacting gas in a $3D$ harmonic trap this
index should be 3.} \label{fig2}
\end{figure}

These results confirm the evolution of the effective dimensionality from $3D$%
, for which $\alpha =3$, to quasi-$2D$ and the effective narrowing of the
trap during condensation.\ \

The situation with an optical lattice (Refs. \cite{lat1} and references
therein) with a period $a_{0}$ inside the trap is more complex. Here one
deals with the Hubbard Hamiltonian, modified by the trap potential, and can
encounter the Mott transition \cite{mott1} which requires full occupancy of
the lattice sites. The latter can occur with lowering of the temperature
when particles gravitate towards the bottom (center) of the trap. With
sufficiently strong on-site repulsion, the localization is practically
inevitable for the condensate in the center of the trap though, of course,
the Mott transition is sensitive to the trap profile \cite{mott1,mott2}. The
increased size of the condensate droplet in comparison to the system without
the lattice changes the normal cloud surrounding the condensate for which it
is possible to disregard the Mott transition.

In the case of low initial density of particles $na_{0}^{3}\ll 1$ and strong
on-site repulsion, the condensation starts at the same temperature $T_{c}$
as without the interaction. The condensate forms in the center of the trap
and rapidly expands in size because of the strong on-site repulsion which
tends to keep the density $n_{c}a_{0}^{3}\approx 1$. As a result, the size
of the condensate droplet $\sigma \sim a_{0}N_{c}^{1/3}$ becomes larger than
$\sigma _{\max }\sim \left( 2\div 5\right) \sigma _{0}$ for traps without
the optical lattice. We will not dwell on potential freezing of the
condensate resulting from the Mott transition and will concentrate on the
condensation of the normal gas outside the condensate droplet.

The main changes are associated with the band nature of the energy spectrum
for particles in the optical lattice and a more complicated form of the wave
functions. For the sake of comparison, in computations we use the same trap
potential and particle scattering length. For the particle effective mass we
use in computations the value \cite{nature1,mott1} $m^{\ast }=16m$.

The single-particle spectrum in the optical lattice $\epsilon \left( \mathbf{%
p}\right) $ has a band structure with a bandwidth $\Delta $. The effect of
the trap potential $U_{tr}\left( r\right) =\frac{1}{2}\hbar \omega \left(
r/\sigma _{0}\right) ^{2}$ on the particles with narrow bands results in
localization of particles with energy $E$ in $2D$ shells $\epsilon \left(
\mathbf{p}\right) +U_{tr}\left( r\right) =E$ of the thickness $\ell \left(
r\right) \sim \left( \Delta /\hbar \omega \right) \sigma _{0}^{2}/r$. [An
exception is the center of the trap, where the gradient of the potential is
small]. The particle wave function consists of three regions: rapid
oscillations within this classically accessible shell and two attenuating
tails beyond the classical turning points. The wave function for a particle
with the energy $E$ decays relatively slowly beyond the turning points,
often as the Airy function of the type $\mathrm{Ai}\left( -\left[ x+\left(
m^{\ast }\Delta /4\hbar ^{2}\nu ^{2}\right) ^{1/3}\right] \right) $ where $x$
is the distance from the "center" of the classically accessible shell for
the particle with the energy $E$ in the direction of the gradient, $\nu =%
\sqrt{2\hbar \omega \left( E-\Delta /2\right) }/\sigma _{0}$, and $m^{\ast }$%
\ is the particle effective mass at the turning point. The spatial
distribution of particles should be calculated taking into account all three
regions since for relatively shallow traps the contribution from the tails
of the wave function can be large. Since such localization suppresses the
accessibility range of narrow-band particles, the density in each point
contains the contributions from the particles in a finite range of energies
that are localized close to this point. For example, since only the
particles with very low energies, $E<\Delta ,$ can reach the center of the
trap, the density in the center is suppressed in comparison with the trap
without the optical lattice inside.

As above, we start from the situation when the particle density above
condensation is low and the (Hubbard) repulsion in the normal phase is
negligible. The condition of low density allows us also to disregard the
Mott transition in the normal phase \cite{mey1}. Since the particles in the
optical lattice are located mostly on the lattice sites of the size $a_{0}$
rather than spread uniformly, the repulsion is more effective than without
the lattice. This means that the size of the condensate droplet $\sigma
\left( T\right) $ should be larger than in the absence of the lattice. This
is illustrated in Figure 3 which presents the ratio $\sigma \left(
T=0\right) /\sigma _{0}$ for identical traps with (curve 1) and without
(curve 2) the optical lattice. The scattering amplitude $a_{s}$, which is
responsible for repulsion, is the same in both cases.

\begin{figure}[ptb]
\includegraphics[angle=270,scale=0.5]{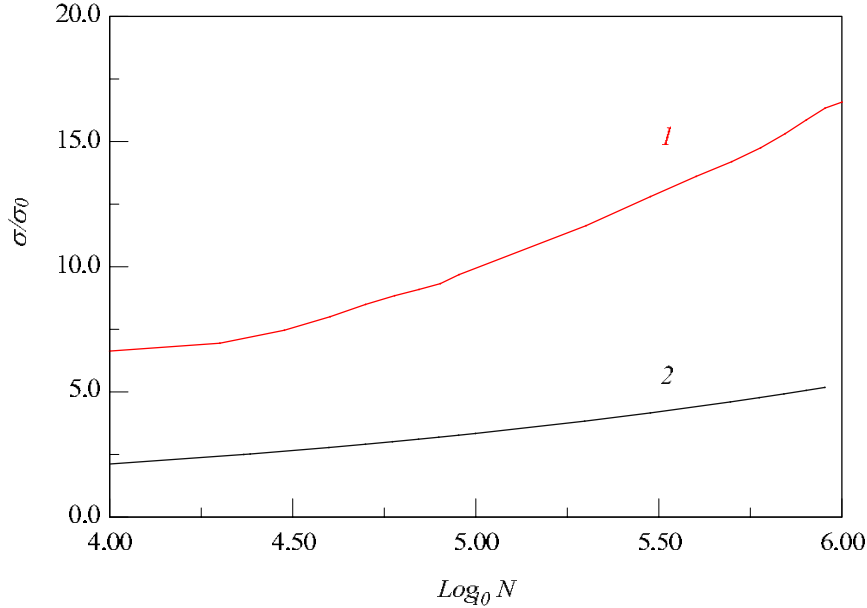}
\caption{(Color online) Size of the condensate droplet $\sigma
\left( T=0\right) $ relative to the size of the trap $\sigma
_{0}$, $\sigma /\sigma _{0}$, with (curve 1) and without (curve 2)
the optical lattice as a function of the number of particles in
the trap. The scattering lengths and effective masses are
identical in both cases. Parameters of the lattice and the trap
are given in the text.} \label{fig3}
\end{figure}

This seemingly innocuous lattice-driven increase in $\sigma $ leads to major
effects and can eliminate a repulsive bump $\left( \ref{eq1}\right) $ in the
center of the trap (at $N_{c}\sigma _{0}^{5}/N_{0}\sigma ^{5}=1$) thus
restoring the potential's parabolic structure in the central area. As a
result, the change in index $\alpha $, Eq. $\left( \ref{eq2}\right) $, is
even more dramatic than the change in $\sigma $, Figure 4.

\begin{figure}[ptb]
\includegraphics[angle=270,scale=0.5]{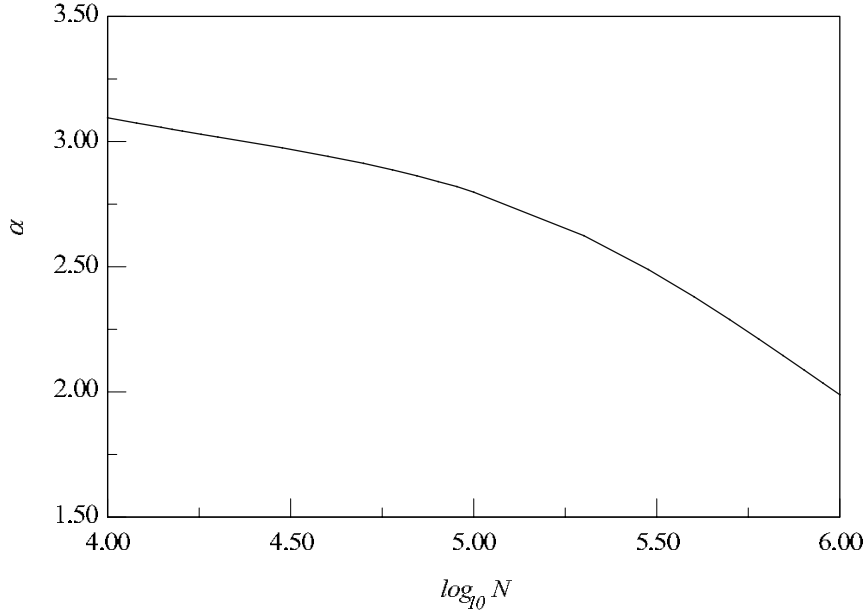}
\caption{Index $\alpha $, Eq.$\left( \ref{eq2}\right) $, as a
function of the number of the trapped particles in the presence of
the optical lattice. Parameters of the lattice used in the
computation are given in the text.} \label{fig4}
\end{figure}

In Fig. 4, $\alpha $ starts from a $3D$ value at small density of particles
which is understandable since there is no repulsive core in the center. With
increasing number of particles the size of the condensate droplet grows
leaving fewer normal particles in the central area and gradually reducing $%
\alpha $ to its quasi-$2D$ value. What is not clear is why does $\alpha $
continue to decline with a further increase in $N$; however, since our
approach loses accuracy beyond $N=10^{6}$, we do not present these data in
the Figure. In general, the decrease in $\alpha \left( N\right) $ is
accompanied by an increase in $T_{c}^{\ast }\left( N\right) $, which in the
presence of the optical lattice grows much faster than $N^{1/3}$-dependence
inherent to a free gas in a trap.

In summary, we calculated the index for a temperature dependence of the
condensate fraction for interacting gas inside harmonic trap. The results
for traps without the optical lattice inside are quite clear: the repulsion
from the condensate droplet pushes normal particles away from the center of
the trap and concentrates them in a relatively thin shell around this
droplet. Then the condensation becomes almost quasi-$2D$ with the index $%
\alpha \approx 2$. The presence of the optical lattice inside the trap
changes the situation. The index $\alpha $ acquires a strong dependence on
the number of particles inside the trap and gradually falls from a $3D$ to a
$2D$ value with an increase in the number of particles. This change in the
index, which is caused by the presence of the optical lattice, is explained
by the wider spread of the condensate droplet and the localization of the
narrow band particles by the trap potential.

\end{document}